\documentclass[a4paper,showpacs,nofootinbib,aps,floatfix]{revtex4}
\def \be {\begin{equation}}
\def \ee {\end{equation}}
\def \bea {\begin{eqnarray}}
\def \eea {\end{eqnarray}}
\input psfig.sty %{epsf}
\usepackage{graphicx}
\usepackage{epsfig}
\usepackage{color}
\usepackage{epsf}
\usepackage{psfrag}

\begin{document}
\title{Constraining thawing and freezing models with cluster number counts}

\author{N. Chandrachani Devi\footnote{E-mail: chandrachani@on.br}}

\author{J. E. Gonzalez\footnote{E-mail: javierernesto@on.br}}

\author{J. S. Alcaniz\footnote{E-mail: alcaniz@on.br}}

\address{Observat\'orio Nacional, 20921-400, Rio de Janeiro - RJ, Brasil}

\date{\today}

\begin{abstract}

Measurements of the cluster abundance as a function of mass and redshift provide an important cosmological test that probe not only the expansion rate but also the growth of perturbations. In this paper we adopt a scalar field scenario which admits both thawing and freezing solutions from an appropriate choice of the model parameters and derived all relevant expressions to calculate the mass function and the cluster number density. We discuss the ability of cluster observations to distinguish between these scalar field behaviors and the standard $\Lambda$CDM scenario by considering the eROSITA and SPT cluster surveys.

\end{abstract}


\maketitle

\section{Introduction}

Nowadays, one of the  most important tasks in Cosmology is to unveil the nature of the mechanism responsible for the current cosmic acceleration. From the observational viewpoint, the cosmological constant $\Lambda$, i.e., the sum over the vacuum energy density of all fields in Nature, remains as the favorite candidate for explaining this phenomenon, but because of its theoretical issues (e.g. fine tuning and cosmic coincidence problems~\citep{Weinberg}), many alternative mechanisms have been proposed (see \citep{rev} for recent reviews). 

Among these, a very light, minimally coupled scalar field $\phi$ whose evolution is driven by the slope of the potential $V(\phi)$ and damped by the cosmic expansion $H$ according to the Klein-Gordon equation $\ddot{\phi} + 3H\dot{\phi} = - dV/d\phi $, is the most extensively studied dark energy case~\citep{quint,Quintessence}. In terms of its potential or still of its equation-of-state (EoS) parameter, $w_\phi(\phi, \dot{\phi}, V)$, these dark energy fields can be broadly classified into two categories: thawing models whose EoS increases from $w_{\phi} \sim -1$, as the field rolls down toward the minimum of its potential with $\dot w_{\phi} > 0$ and cooling scenarios in which an initial $w_{\phi} > -1$ EoS decreases to more negative values with $\dot w_{\phi} < 0$. A special case of the latter is the so-called freezing models, in  which the potential has a minimum at $\phi = \infty$~\citep{cl}. Examples of thawing-type scenarios are a pseudo Nambu-Goldstone boson $V(\phi) = M^4\cos^2(\beta \phi)$~\cite{ioav}
 and exponential potentials for dilaton fields $V(\phi) = M^4\exp(-\alpha\phi)$~\cite{exp} (see also \cite{ioav}) whereas the freezing behavior is found in models of the type $V(\phi) = M^{4 + n}\phi^{-n}$ and $V(\phi) = M^{4 + n}\phi^{-n}\exp(\beta \phi^2)$ for $n >0$~\cite{freezing}.

Distinguishing among these (and other) physical features of the dark energy potential from cosmological data constitutes an important approach to the cosmic acceleration problem, not only because it may indicate possible routes of solutions but also because it will reduce considerably the range of possibilities. In this regard, measurements of the cluster abundance as a function of mass and redshift are coming up with the potential to improve current constraints on cosmological parameters \cite{cluster,cluster2}, including the dark energy EoS, the rms mass fluctuations,  
the matter density parameter, the total neutrino masses, etc. These data have the  advantage of probing both the expansion rate and the growth of perturbations, thereby being complementary to other cosmological probes such as the Comic Microwave background (CMB) anisotropies, observations of type Ia supernovae and measurements of baryon acoustic oscillations (BAO). 

In what follows, we investigate to what extent galaxy cluster number counts can be used to distinguish the thawing and freezing behaviors discussed above (see also Sec. II) as well as between these dynamical scenarios and the standard $\Lambda$CDM model. In our analysis, we consider the scalar field model discussed in Ref.~\cite{Carvalho2006} which admits both solutions from an appropriate choice of the model parameters. Keeping the range of cosmological parameters within the allowed region by the current observations and considering two specific cluster surveys, namely eROSITA and SPT, we calculate the number density of cluster and the redshift distribution of cluster number counts for this scenario and discuss the ability of these observations to distinguish between the thawing and freezing behaviors.

\section{Background Cosmology}

We consider a flat Friedmann-Lama\^itre-Robertson-Walker universe driven by a non-relativistic matter which clusters under the action of gravity and a scalar field responsible for the current cosmic acceleration whose conservation equation takes the form
\begin{equation}
\dot{\rho}_{\phi}+3H(\rho_{\phi}+p_{\phi}) = 0\;,
\label{conservation}
\end{equation}
where 
\begin{equation} \label{pr}
p_\phi = \frac{1}{2}\dot{\phi}^2 - V(\phi)\; \quad \mbox{and} \quad \; \rho_\phi = \frac{1}{2}\dot{\phi}^2 + V(\phi)\;,
\end{equation}
are, respectively, the scalar field pressure and energy density and a dot denotes derivative with respect to time. The above equation can also be written as
\begin{eqnarray} 
\frac{\partial \phi}{\partial a} = \sqrt{-\frac{1}{8\pi a G \rho_\phi}\frac{\partial \rho_\phi}{\partial a}}\;,
\end{eqnarray}
where $a$ is the cosmological scale factor. In Ref.~\cite{Carvalho2006}, the following \emph{ansatz} on the scale factor derivative of the energy density was introduced:
\begin{equation} 
\label{ansatz}
\frac{1}{\rho_{\phi}}\frac{\partial \rho_{\phi}}{\partial a} = -\frac{\lambda}{a^{1-2\alpha}}\;,
\end{equation}
where $\alpha$ and $\lambda$ are arbitrary parameters. When combined with the definitions of $p_\phi$ and $\rho_\phi$ mentioned above, Eqs. (\ref{conservation})-(\ref{ansatz}) provide (for details, see~\cite{Carvalho2006})
\be
\label{gpotential} V(\phi)= f(\alpha, \phi)
\rho_{\phi,0}\exp\left[-\lambda\sqrt{\beta}\left(\phi + {\alpha
\sqrt{\beta} \over 2} \phi^2 \right)  \right], \ee
with
$f(\alpha, \phi) =
[1-{\lambda\over6}(1+\alpha\sqrt{\beta}\phi)^2]$ and $\beta = 8\pi G/\lambda$. It has been pointed out by the previous studies of this model that in the limit of $\alpha \rightarrow 0$ the potential (\ref{gpotential}) reproduces the exponential potential studied in Refs. \cite{exp,RatraPeebles} whereas for all values of $\alpha \neq 0$, it is dominated by the quadratic contribution $\phi^2$, admitting a wider range of solutions. Once the form of potential is obtained, one can easily check the behavior of field evolution through the equation of state parameter
\be
\label{wa}
w_{\phi}(a) = -1 + \frac{\lambda}{3}a^{2\alpha}\;.
\ee
An important point about the above \emph{ansatz} (\ref{ansatz}) is that with an appropriate choice of the parameters $\lambda$ and $\alpha$, one can obtain thawing and freezing behaviors for the EoS parameter above, which will be essential for the cluster analysis discussed in Sec. IV.

\begin{figure*}
\psfig{figure=conNew.eps,width=3.1truein,height=2.3truein}
\hspace{0.7cm}
\psfig{figure=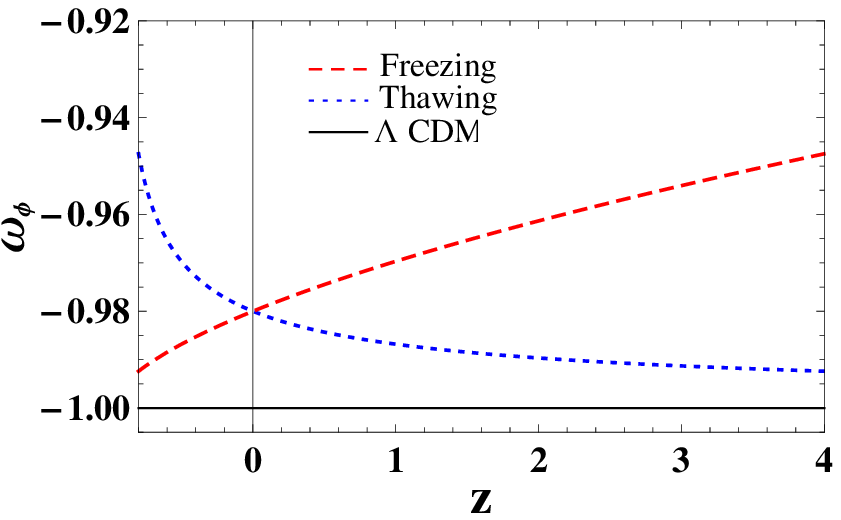,width=3.1truein,height=2.3truein}
\caption{{\it{Left:}} Countours in the $\alpha - \lambda$ plane from a joint analysis involving SNe Ia + CMB + BAO data. {\it{Right:}} Evolution of the equation of state parameter $w_{\phi}$ from Eq. (6). Short and long dashed curves correspond to thawing and freezing models, respectively, with $\alpha = \pm 0.3$ and $\lambda = 0.06$. The solid curve stands for the standard $\Lambda$CDM model.}
\label{fig:wz}
\end{figure*} 

\subsection{Observational Analysis}

In order to obtain the allowed range of the parameters $\alpha$ and $\lambda$, we perform a statistical analysis using recent cosmological data from type Ia supernova (SNe Ia)~\cite{SN}. We also use Cosmic Microwave Background (CMB)~\cite{CMBplank,shafer} and Large-Scale Structure (LSS) data~\cite{BAOa,BAO} to help break the degeneracy between the scalar field parameters and the matter density parameter $\Omega_{m0}$. For SNe Ia data, we use the publicly available latest Union2.1 compilation~\cite{SN} which consists of 580 data points. For the CMB, we use only the measurement of the CMB shift parameter~\cite{bond}
$$
{\cal{R}} = \sqrt{\Omega_{m0}}\int_{0}^{z_{ls}}\frac{dz'}{H(z')} = 1.7407 \pm 0.0094\;,
$$ 
as measured by the Plank collaboration \cite{CMBplank,shafer}, where $z_{ls} = 1091$. The LSS information we use is the latest Baryonic Acoustic Oscillation (BAO) data set listed out in table III of Ref.~\cite{BAO} which comprises distance measurements at six different redshifts from SDSS, WiggleZ and 6dFGS redshift surveys. We combine the above probes to provide constraints on our model parameters by using a joint $\chi^2$ analysis with  $\chi(p)^2_{tot}=\chi^2_{SNIa}+\chi_{cmb}^2+\chi_{BAO}^2$ (we refer the reader to Ref.~\cite{test} for details on the statistical analysis).

For a minimally coupled scalar field model with a canonical kinetic term [see Eq. (\ref{pr})], the equation-of-state parameter $w_{\phi} \equiv p_{\phi}/\rho_{\phi}$ lies necessarily in the interval [-1,1] (Note also that for  negative values of $\lambda$ the potential $V(\phi)$ becomes a complex quantity). Therefore, we consider values of $\lambda$ in the interval of [0, 1]. Our joint analysis of SNe Ia + CMB + BAO data results into best-fit values of $\Omega_{m0}=0.29 \pm 0.01$, $\lambda  = 0.00 \pm 0.06$ and $\alpha = 0.30 \pm 1.20$, which is in full agreement with the standard $\Lambda$CDM model (see Fig. 1a).  However, since our main goal is to discuss the influence of time evolving scalar field models on cluster counts, we will consider the 1$\sigma$ interval for $\lambda$. Therefore, without loss of generality to the subsequent analyses, from now on we particularize our study to the values of $\alpha$ within the interval [-0.3, 0.3] and $\lambda = 0.06$.

Fig. (\ref{fig:wz}b) shows the evolution of $w_{\phi}(a)$ for some specific values of $\alpha$ and $\lambda = 0.06$. From the figure, one can easily observe that values of $\alpha > 0$ and $\alpha < 0$ correspond, respectively, to thawing and freezing behaviors of the scalar field model. It is worth mentioning that in the case of an increasing EoS parameter (thawing), the current cosmic acceleration is a transient phenomenon with the scalar field $\phi$ leading the Universe to a future decelerated phase whereas in freezing scenarios, the field drives the Universe to an eternal quasi-de Sitter phase.

\section{Halo Abundances}

Assuming that cluster of galaxies are surrounded by a cold dark matter (CDM) halo we perform our analysis considering the abundance of CDM halos instead of cluster of galaxies. This CDM halo abundance is widely studied either through N-body simulation~\cite{Vikhlinin,Bhattacharya2011,Courtin2011} or through a semi-analytical approach~\cite{Gunn,wang, spherical1,spherical2}. Since our primary goal is to check to which level one can distinguish among the above mentioned scalar field models through the cluster number density, we adopt the semi-analytical approach of the spherical collapse model for our study of the structure formation. 
 
\subsection{Non-linear and Linear Matter Evolution}

The main ingredients for studying the cluster number density are the critical density contrast $\delta_{c}(z)$ at the collapse point, above which structure collapses, and the growth factor, $D(a)=\delta(z)/\delta(0)$. In order to calculate these quantities, we assume that the dark matter and dark energy components are uncoupled (except gravitationally) and study the non-linear evolution of matter perturbation by adopting a spherically over-dense region of radius $r$ numerically. This over-dense region follows the Raychaudhuri equation;
\begin {equation} 
\frac{\ddot{r}}{r}=-4 \pi G\left[ \left(w_{\phi}(r)+\frac{1}{3}\right) {\rho_{\phi cl}}+\frac{1}{3}{\rho_{m cl}}\right] \;,
\label{eq:raychaudh}
\end{equation}
where $\rho_{\phi cl}$ and $\rho_{m cl}$ are, respectively, the dark energy and dark matter densities of the cluster.  However, on scales smaller than the horizon, one can assume that the dark energy component is smooth and homogeneous, so that $\rho_{\phi} = \rho_{\phi cl}$. Therefore, we study fluctuations on the cold dark matter only. 

\begin{figure*}
\label{fig:dencontrast}
\psfig{figure=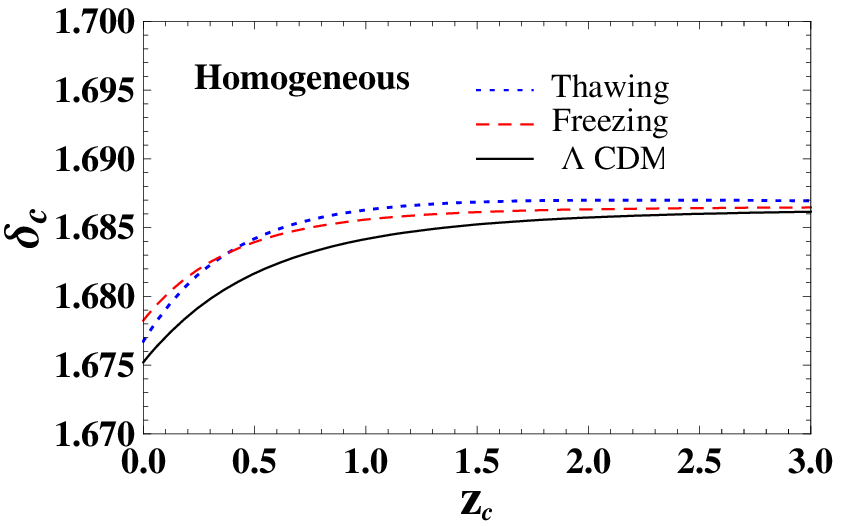,width= 2.8truein,height=2.3truein}
\hspace{0.7cm}
\psfig{figure=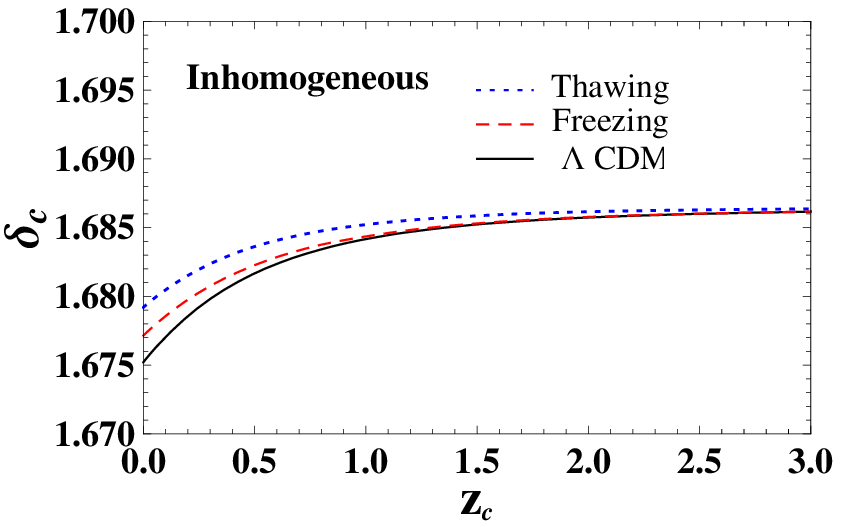,width=2.8truein,height=2.2truein}
\hskip 0.2in
\caption{The linearly extrapolated density contrast $\delta_c$ at the collapse point as a function the collapse redshift $z_c$ for the models considered, i.e., freezing, $\alpha =-0.3$ and thawing, $\alpha =+0.3$ with the $\Omega_{m0} =0.29$. The $\Lambda$CDM case is shown for comparison. {{Left}} and {{right}} panels correspond, respectively, to homogeneous and inhomogeneous cases discussed in the text [see Eqs. (\ref{eq:perturbation}) and (\ref{eq:I})]. }
\end{figure*} 

Following \cite{wang,spherical1,spherical2}, we normalize the Friedmann equation
%(\ref{eq.fried}) 
along with Eq.(\ref{eq:raychaudh}) at the turn around time after introducing the variables $x=\frac{a}{a_{t}}$ and $y=\frac{r}{r_t}$, i.e.,
\begin{equation}
  \dot{x}^2={H_t}^2\Omega_{m,t}[ \Omega_m(x)x]^{-1}
\label{eq:background} 
\end{equation}
 and
\begin{equation}
{\ddot y}=-\frac{H_{\rm t}^{2}\Omega_{m,t}}{2}
\left[ \frac{\zeta}{y^{2}}+\frac{1-\Omega_{m,t}}{\Omega_{m,t}} y g(x,y)\right] \;\;\;
\label{eq:perturbation}
\end{equation} 
with the function of $g(x,y)$
\begin{equation}
g(x,y)=\left\{ \begin{array}{cc}
       \displaystyle\left[1+3w(r(y))\right]\frac{f(r(y))}{f(a_t)} &
       \mbox{Inhomogeneous}\\
       \left[1+3w(x)\right]f(x) & \mbox{Homogeneous}
       \end{array}
        \right.
\label{eq:I}
\end{equation}
and the dark energy density function is rewritten as $\rho_{\phi}=\rho_{\phi0}{f(a)}$ with  ${f(a)}= \exp\left[3\int_{a}^{1}\left(\frac{{1+w(u)}}{u}\right) {\rm d}u\right]. $ 
Here 
$\zeta(z)=(\rho_{cl}/\rho_{b})|_{x=1}$ corresponds to the overdensity at turn-around time $t$. We determine $\zeta(z)$ by solving (\ref{eq:background}) and (\ref{eq:perturbation}) simultaneously using the  boundary conditions $dy/dx|_{x=1} = 0, y|_{x=1} = 1$ and $y|_{x=0}=0$. Once $\zeta$ is known, we can evaluate the linear density contrast at the time of collapse point $\delta_c$ as a function of redshift using (\ref{eq:background}), (\ref{eq:perturbation}) and the linear growth factor, $D(z)$:
\bea
\delta_c = \left[\left(\frac{\rho_{mcl}}{\rho_{m}}-1\right) \frac{1}{D(a)_{a\to 0}} \right]_{a \to 0}D(a) \\\nonumber
= \left[\left(\frac{x}{y}\right)^3\zeta-1\right]_{x\to 0}\frac{D(x)}{D(x)_{x\to 0}}.
\eea
Here, the linear growth factor $D(z)$ is obtained by solving the linearised evolution equation of matter Eq (\ref{eq:raychaudh}), written in term of $\delta =\frac{\rho_{mcl}-\rho_{m}}{\rho_{mcl}}$ as
\begin{equation}
{\ddot{\delta}}  +  2\frac{\dot{a}}{a}{\dot{\delta}}
  = 4\pi G \rho_{m}\delta 
  = \frac{3}{2}H_{0}^2\Omega_{m0}a^{-3}\delta \;,
\label{eq:lineardensity}
\end{equation}
with the initial conditions, $\delta(a)\to a$ and $d\delta/da \to 1$ at $a \to a_i =10^{-3}$ (for details we refer the reader to~\citep{wang,spherical1}).
Another approach to calculate the extrapolated linear density contrast $\delta_c$ as a function of collapse redshift has been discussed in \cite{Campanelli}. Both approaches are found to be in agreement with each other.  In Fig. 2, we depict the linearly extrapolated density contrast at the collapse point $\delta_c(z)$ as a function of collapse redshift for different time-evolution of the EoS (\ref{wa}). We consider values of $\alpha=-0.3$ (freezing) and $\alpha =+0.3$ (thawing) and also show the well-known $\Lambda$CDM prediction (solid black line). As expected, all the models considered asymptotically approach to the Einstein-de Sitter limit at high-$z$.  For completeness, we also examine the case where the dark energy field is clustered (see, e.g., Refs. \cite{Basilakos,Basilakos2009}). From the previous results of $\delta_c$, we found that the homogeneous case shows slightly higher value than the inhomogeneous one in both freezing and thawing models. As discussed in \cite{spherical1}, this is an expected 
result since in the latter case there is an extra repulsive effect inside the cluster due to inhomogeneous dark energy which leads to lowering the linear density contrast.

\begin{figure*}
\label{fig:dndlnmhomo}
\psfig{figure=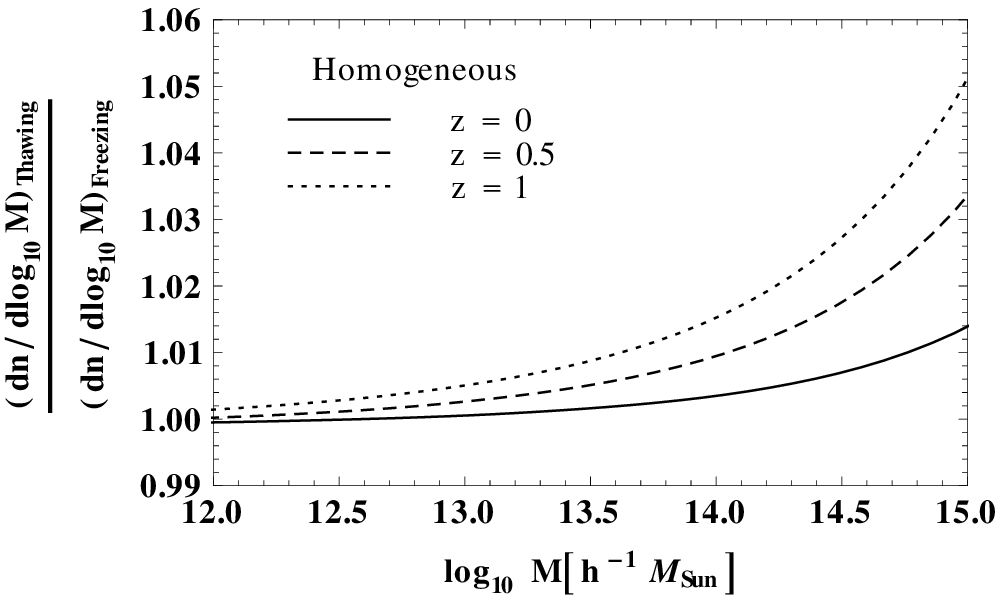,width= 2.32truein,height=2.4truein}
\psfig{figure=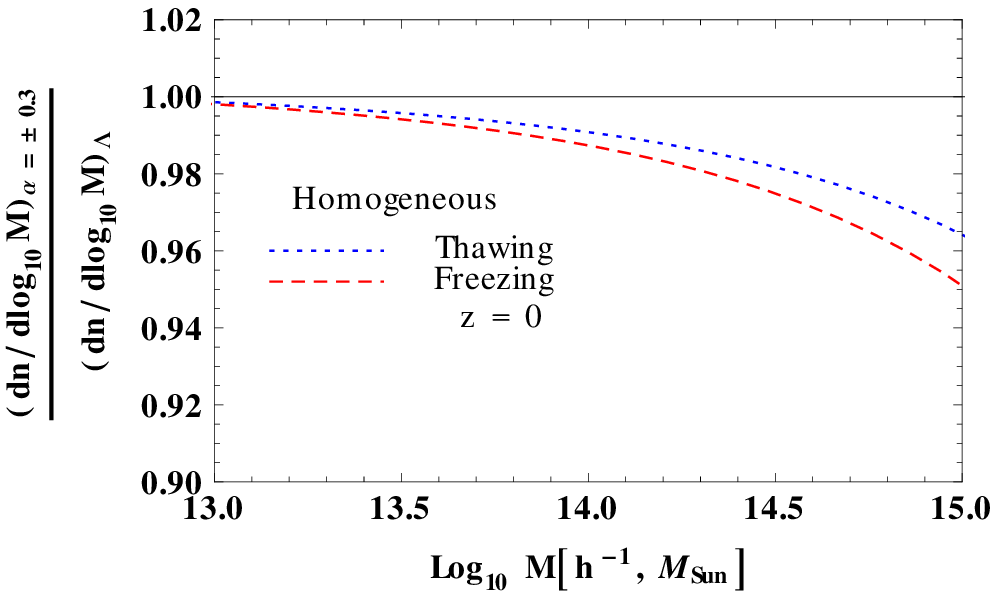,width=2.32truein,height=2.4truein}
\psfig{figure=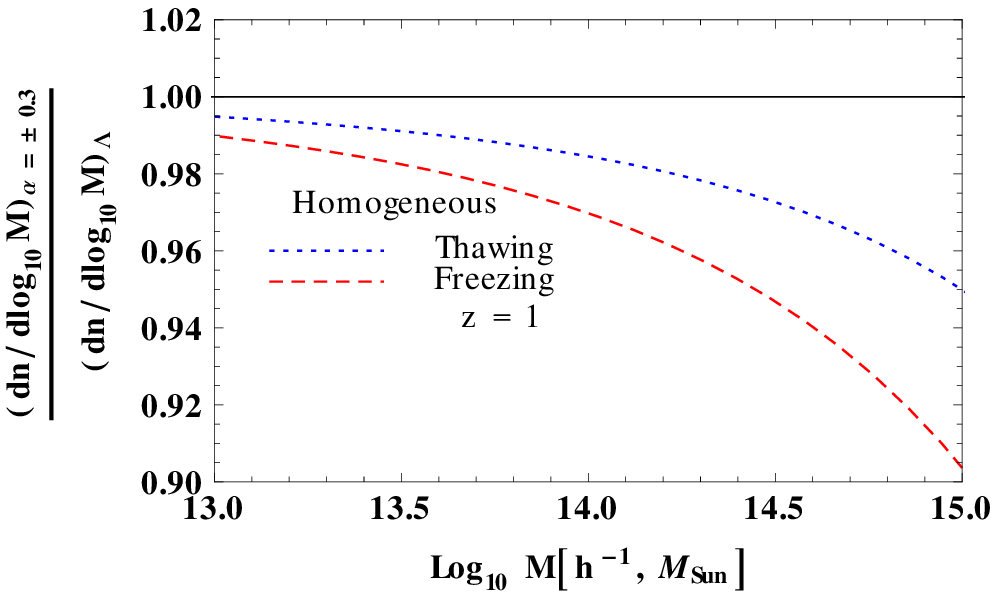,width=2.32truein,height=2.4truein}
\psfig{figure=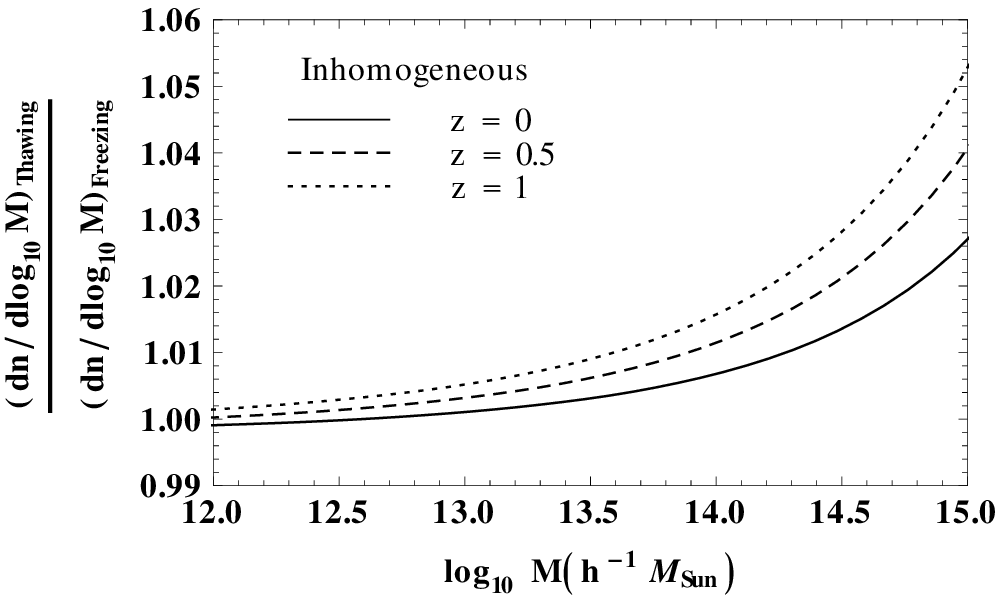,width= 2.32truein,height=2.4truein}
\psfig{figure=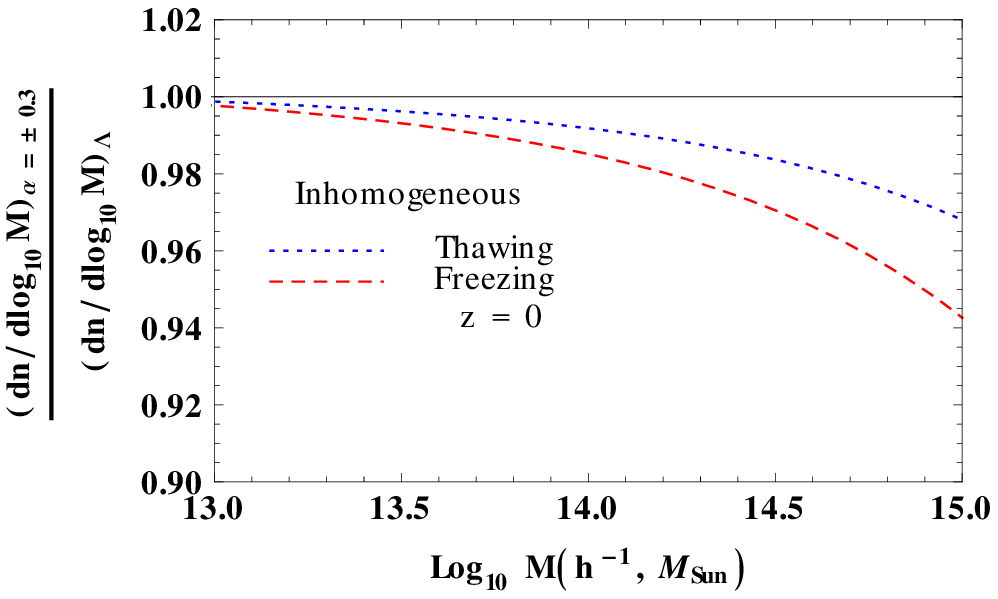,width=2.32truein,height=2.4truein}
\psfig{figure=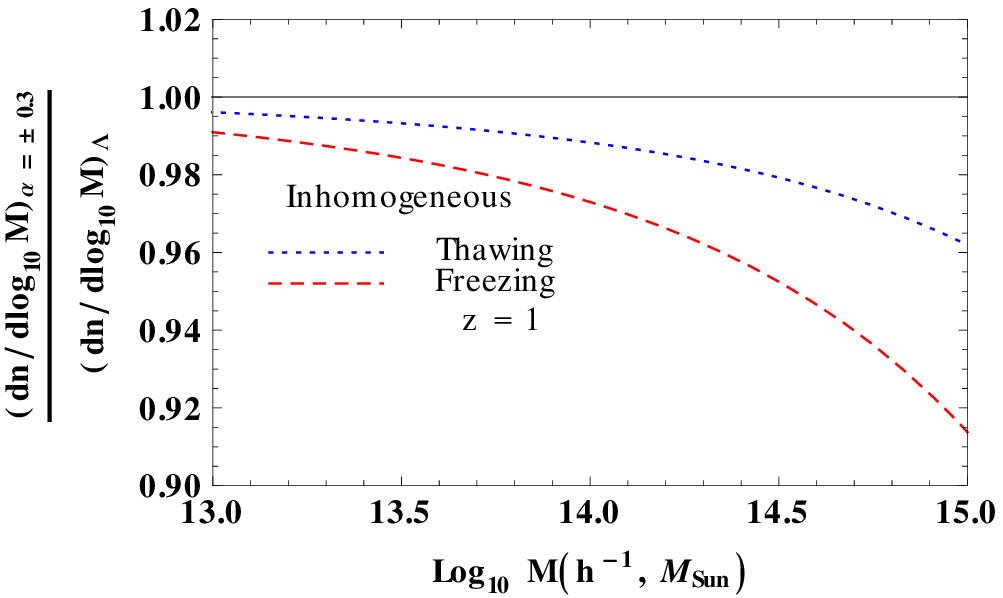,width=2.32truein,height=2.4truein}
\hskip 0.2in
\caption{{\it Left panels}: The difference in the comoving number density between the thawing and freezing dark energy models as function of the cluster mass $M$ at some particular redshifts. {\it{Middle {\rm{and}}  Right panels}}: The difference in the comoving number density as function of the cluster mass $M$ between both thawing ($\alpha =+0.3$) and freezing ($\alpha = -0.3$) models with that of the concordance $\Lambda$CDM model.}
\end{figure*}

\subsection{Cluster Number Counts}
Cluster number count is considered to be the most immediate effect on cosmic structure when different cosmological models are considered. 
In what follows, we calculate the comoving number density of collapsed objects within a  mass range between $M$ and $M+dM$ at a particular redshift $z$, given by 

\begin{equation}
\frac{dn}{dM}(M,z) = \frac{{\rho}_{m}}{M}\frac{d \ln \sigma^{-1}(M,z)}{dM}f(\sigma)\;,
\label{eq:denfunction}
\end{equation}
where  ${\rho}_{m}$ is the comoving background density and $f(\sigma)$ is the mass function. Recently, a large number of analysis have focused on developing  fitting formulas of $f(\sigma)$ based on fits to simulated data (see, e.g., \cite{tinker,jenkins,Reed,warren,manera,crocce}). However, they differ mainly on the high mass end. In this work, we adopt a modified form of Press-Schechter \cite{PS} mass function $f(\sigma)$ proposed by Sheth and Tormen in Ref. \cite{ST}, which is widely studied and reasonably in agreement with the numerical simulations, i.e.,
\begin{equation}
\label{st}
f(\sigma) = A\sqrt{\frac{2a_1}{\pi}}\left[1+\left(\frac{\sigma^2}{a_1\delta_c^2(z)}\right)^p\right]\frac{\delta_c(z)}{\sigma}\exp\left[- \frac{\delta_c^2(z)a_1}{2\sigma^2}\right]\; ,
\end{equation}
with $ A=0.322, a_1=0.707$ and $p=0.3$. $\sigma(z,R)$ represents the mass variance of the linear density field smoothed out by a top hat filter on a comoving length scale $R$ and extrapolated to redshift $z$ where the halos are identified. It is given by
\begin{equation}
\sigma^2(z,R) = \frac{1}{2\pi^2}\int_0^{\infty} k^3 P_{m}(k,z)W^2(kR)\frac{dk}{k},
\end{equation}
where  $W(kR) = 3 \left(\sin(kR)-kR\cos(kR)\right)/(kR)^3$ 
is the Fourier transform of spherical top-hat filter. $P_m(k,z) = Ak^{n_s}T^2(k)D^2(z)$ represents the matter power spectrum where a standard power law of the primordial density fluctuation is being used with its overall normalization factor $A$ and tilt $n_s$, respectively. The growth factor, $D(z)$,  obtained from Eq.(\ref{eq:lineardensity}), is normalized at the present epoch. 
For the transfer function,
$T(k)$, we use Eisenstein and Hu transfer function~\cite{Hu}.
Generally, the amplitude of the matter power spectrum is normalised using the mass variance calculated today on a scale of  $8{\rm Mpc h}^{-1}$. However, we follow the normalization procedure of \cite{Basilakos} in which the WMAP results are used \cite{WMAP7,WMAP9}.  Since we are mainly interested in differences of the predicted cluster number counts from thawing and freezing models, we put a flat prior on the cosmological 
parameters according to WMAP-9+BAO+H0 best fit values~\cite{WMAP9}, i.e., $H_{0} = 69.33~{\rm Km/s/Mpc},~\Omega_{B0}h^2 = 0.02266, n_s= 0.971~{\rm and}~\sigma_{8,\Lambda} =0. 83 ~{\rm with}~\Omega_{_{CDM}}h^2 = 0.1157$ to obtain the results for cluster number density.
Once the number density per unit mass $dn(M,z)/dM$ is known, it is easy to obtain the number of clusters per redshift interval $dz$ above a given minimum (threshold) mass $M= M_{min}$ as 
\be
\frac{dN}{dz}= f_{sky} \frac{dV}{dz}\int_{M_{min}}^{\infty} \frac{\rho_{m}}{M}\frac{d \ln \sigma^{-1}(M,z)}{dM}f(\sigma; ST) dM.
\label{eq:denfunctiondz}
\ee
where $f_{sky}$ is the observed sky fraction and $\frac{dV}{dz}$ is the comoving volume element which depends on the background cosmology.
%\be 
%\frac{dV}{dz}= \frac{4\pi}{H(z)} \left[\int_0^z \frac{dz^{'}}{H(z^{'})}\right]^2
%\ee
%and $r(z)$ is the comoving radial distance at redshift $z$ 
%\be 
%r(z)= c\int_0^z \frac{dz^{'}}{H(z^{'})}.
%\ee
The effect of cosmology enters through the mass function and the volume element whereas the minimum mass limit, $M_{min}$, and the fraction of sky, $f_{sky}$, will be set according to the cluster surveys. 
The abundance of galaxy clusters is also sensitive to the
matter density, $\Omega_{m0}$, and to the amplitude of density fluctuations, $\sigma_{8}$. In the next section, we discuss the results obtained from the studies of the cluster number density and the redshift distribution of cluster number counts per unit redshift for the thawing and freezing behaviors of the scalar field model discussed in Sec. II.

\begin{figure}
\psfig{figure=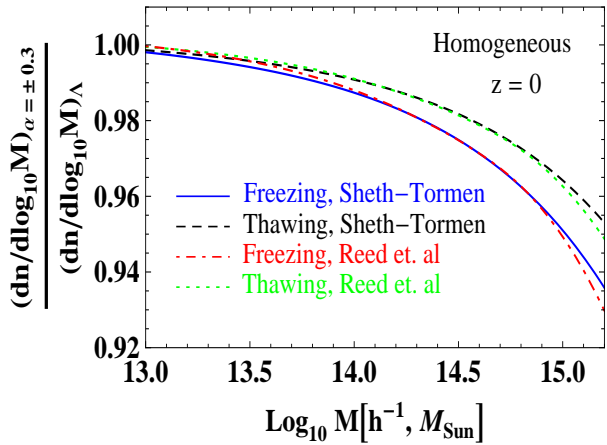,width=3.1truein,height=2.3truein}
\caption{The difference in the comoving number density as function of the cluster mass $M$ between both thawing ($\alpha =+0.3$) and freezing ($\alpha = -0.3$) models with that of the concordance $\Lambda$CDM model, considering two different mass functions, namely the Sheth-Tormen  mass function (\ref{st}) and mass function introduced by Reed {\it{et. al}} \cite{Reed}.}
\label{fig:reed}
\end{figure}

\section{Discussion and Results}

In order to check at which level one can distinguish between these two scalar field behaviors, we plot the predicted difference in the cluster number density as a function of the cluster mass $M$
at redshifts $z = 0, 0.5$ and 1 (left panels of Fig. 3). Clearly, this difference becomes more significative at higher-$z$ and for more massive clusters. 
At $M =10^{13.5}{\rm M_{\odot} h^{-1}}$, for instance, {in case of homogeneous dark energy} it varies from  {0.2\%} up to {0.8\%} between $z = 0$ and 1, whereas for the same redshift interval and $M =10^{14.5}{\rm M_{\odot} h^{-1}}$, the predicted difference between thawing and freezing models varies from {0.7\% to 2.6\%}. 

In Fig. 3 (middle and right panels), we compare the cluster number density of both freezing and thawing scalar fields with the one predicted by the standard $\Lambda$CDM model for both homogeneous and inhomogeneous cases. For lower mass clusters, the three behaviors are indistinguishable at $z = 0$ (upper middle panel) and very similar at $z = 1$ (upper right panel). A clear distinction in behavior is seen for more massive clusters. At $z = 1$, the difference between the predictions of freezing scenarios and the standard model considering the homogeneous case is {$\simeq 3\%$ } for $M =10^{14}{\rm M_{\odot} h^{-1}}$ and only {$\simeq 1.5\%$ } between thawing scenarios and the $\Lambda$CDM model. Such a result can be understood in terms of the behavior of the EoS for thawing models which evolve more closely to $w = -1$ than freezing scenarios (see Fig. 1b). The results for the inhomogeneous dark energy are shown in the lower middle and right panels of the Fig. 3. We found a slightly larger difference between 
freezing and thawing models  than in the previous case.

In order to test the dependence of these results with the Sheth-Torman mass function [Eq. (\ref{st})], we repeat our calculations using the mass function introduced in Ref.~\cite{Reed}. A comparative result considering only the homogeneous case and $z = 0$ is shown in Fig. 4, where we observe a slight difference in number density at the high mass limit. Although not shown  in the figure, a similar conclusion is also drawn for the inhomogeneous and higher-$z$ cases.

From Eq. (14), we can also predict the cluster number distributions expected to be detected in a particular cluster survey. Clusters are detected through the Sunyev-Zel'dovich effect (SZ) and  X-ray flux surveys, via weak or strong lensing surveys and optical surveys. To relate the survey properties to the extent in mass and redshift space of the resulting cluster catalogue, it is necessary to link the mass and redshift  of an individual cluster to the relevant observable, namely X-ray flux or SZ flux. This is done through realistic scaling relations, which depends on the cluster physics as well as on the nature of surveys (see \cite{massca,survey_mass} for a detailed discussion), since the limiting mass $M_{\rm min}(z)$ of any survey depends on the limiting flux of that particular survey. Here, we follow the method described in \cite{Basilakos,massca} for converting the limiting fluxs to limiting halo masses $M_{\rm min}(z)$ for the eROSITA\footnote{http://www.mpe.mpg.de/eROSITA} and SPT\footnote{http://
pole.uchicago.edu/} surveys. The extended Roentgen Survey with an Imaging Array 
(eROSITA) is an upcoming X-ray survey with a sky coverage of $f_{sky} \simeq 0.485$ and limiting flux of $F_{lim} =3.3\times 10^{-14} {\rm erg s^{-1} cm^{-2}}$ in the energy band [0.5-2.0] KeV. The South Pole Telescope (SPT) SZ survey is an ongoing survey which has a sky coverage of $\sim 4000$ deg$^2$ (corresponding to a fraction $f_{\rm sky} \approx 0.097$) with a limiting flux density $f_{\nu_0, {\rm lim}} = 5$ mJy at the frequency $\nu_0 = 150$ GHz.

\begin{figure*}[t]
\label{fig:dndlnMdiff}
\psfrag{z}[b][c][1][0]{{\bf { $z$}}}
\psfrag{dNe}[b][c][1][0]{{\bf { $\frac{dN}{dz}\times 10^4 (M > M_{min})$}}}
\psfrag{dNs}[b][c][1][0]{{\bf { $\frac{dN}{dz}\times 10^3 (M > M_{min})$}}}
\psfig{figure=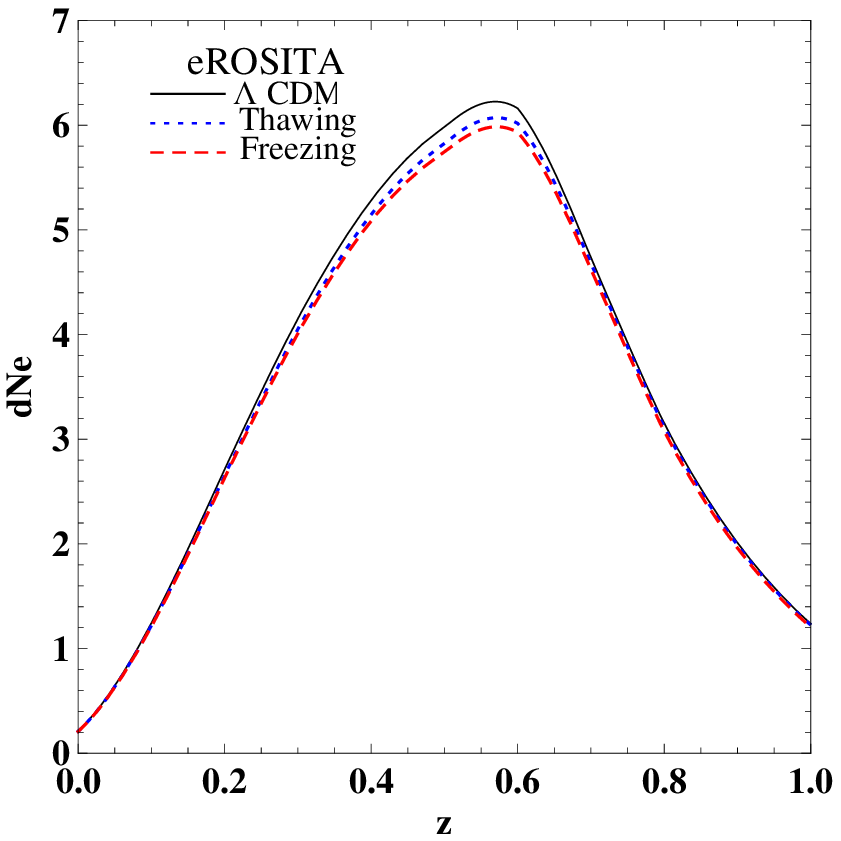,width=2.6truein,height=2.45truein}
\hskip 0.3in
\psfig{figure=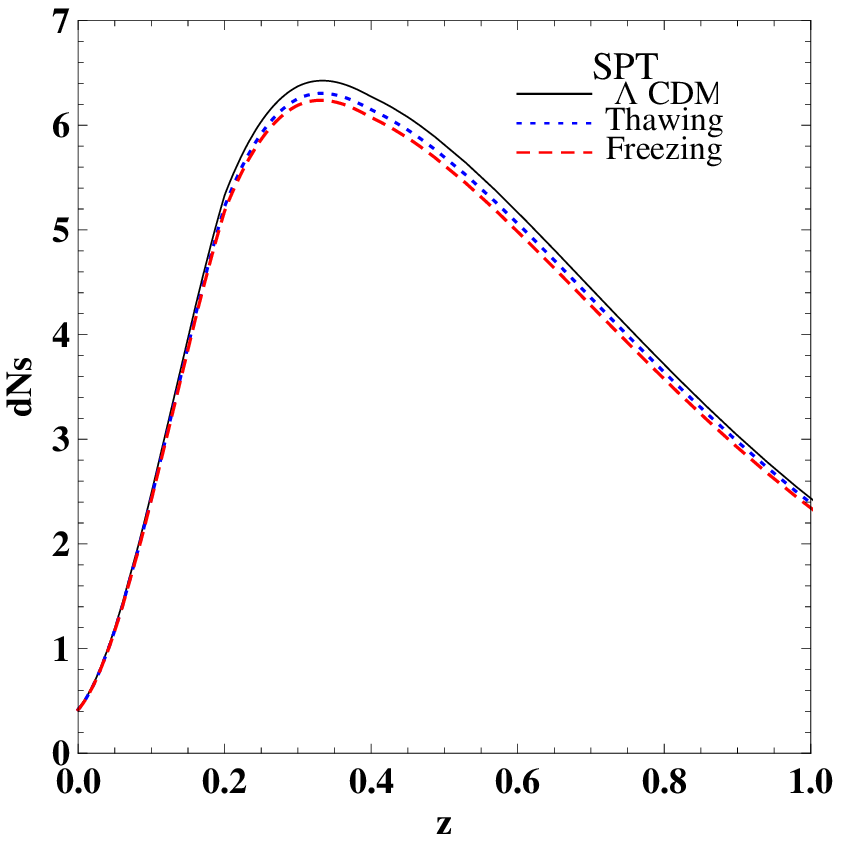,width=2.6truein,height=2.45truein}
\caption{Expected redshift distribution in the cluster number counts of the thawing ($\alpha = +0.3$) and freezing ($\alpha = -0.3$) dark energy models for the mentioned surveys (see discussion in the text) for homogeneous dark energy case. The $\Lambda$CDM model prediction is also shown for comparison. }
\end{figure*}

Figure 5 shows the expected redshift distribution of clusters with the limiting halo masses, $M_{\rm min}(z)= {\rm Max}[10^{14} h^{-1} M_{\odot},M_{cal}(z)]$ where $M_{cal}(z)$ is the mass calibrated using the cluster scaling relations described in \cite{Basilakos,massca}. For numerical simplification, we place the maximum mass limit of $ M_{max} = 10^{16} {\rm h^{-1} M_{\odot}}$. Considering the eROSITA specifications above we find a difference of $\sim 850$ clusters at $z \simeq 0.6$ between thawing and freezing models, which is around two times larger than the estimated eROSITA uncertainty, $\Delta N \simeq 500$ clusters \cite{erosita}. This clearly points to a real possibility of distinguishing between these two behaviors with current planned surveys. Another interesting result arises when we compare these scalar field models with the standard $\Lambda$CDM case. We find a difference in the number of clusters of $\sim 1600$ and $\sim 4500$ between thawing/$\Lambda$CDM and freezing/$\Lambda$CDM, respectively, 
which is $\simeq 3 - 5$ times larger the survey precision.

In the case of SPT, we find that the difference between the theoretical predictions of these two scalar field scenarios is $\sim 80$ clusters at $z \simeq 0.5$ which is within the limit of the survey precision, i.e., $\Delta N \simeq 100-150$ clusters \cite{spt}. When compared with the $\Lambda$CDM model, the largest predicted difference in cluster number counts is $\sim 120$ and $\sim 200$ at $z \simeq 0.4$ for the thawing and freezing scenarios, respectively.

\section{Conclusions}

Thawing and freezing scalar fields are realistic candidates for dark energy. In terms of the field dynamics, while the former has
been frozen until the onset of cosmic acceleration, when it starts to roll down towards the minimum of its potential $V_{min}(\phi)$, the latter was already rolling to $V_{min}(\phi)$ at that time and dominates the Universe as it slows down. From the theoretical viewpoint, both behaviors have important implications for the cosmic evolution, with thawing potentials driving the Universe to a transient accelerating phase while freezing models to an eternal quasi-de Sitter expansion. 

In this paper we have adopted a specific scalar field model (Eq. 4), which admits both thawing and freezing solutions with an appropriate choice of the parameter $\alpha$, and discussed the possibility of current and planned survey distinguishing these two behaviors through observations of cluster number counts. Adopting the spherical collapse model, we have studied the structure formation in these scenarios and derived all relevant expressions to calculate the mass function and the cluster number density using the Sheth-Torman formalism. We have found that the cluster counts predictions of the $\Lambda$CDM and thawing models are very similar, which make this latter class more difficult to be distinguished than freezing ones.
 
By considering eROSITA and SPT surveys, we have also calculated the theoretically expected cluster redshift distribution for cluster halos size 
(Fig. 4). From our analysis, we have found that, while SPT would not be able to distinguish between thawing and freezing models, the more precise eROSITA observations have the potential not only to discriminate these two scalar field behaviors but also to differentiate them from the standard $\Lambda$CDM model.

\begin{acknowledgments}
The authors wish to thank T. Roy Choudhury for helpful discussions. This work is supported by Conselho Nacional de Desenvolvimento Cient\'ifico e Tecnol\'ogico (CNPq), FAPERJ and INEspa\c{c}o. 
\end{acknowledgments}

\end{document}